\documentclass[12pt]{article}
\usepackage{amssymb,amsmath,epsfig}

\begin{document}

\title{\bf Phantom Accretion onto the Schwarzschild de-Sitter Black Hole
\footnote{Supported by the Higher Education Commission, Islamabad,
Pakistan through the Indigenous Ph.D. 5000 Fellowship Program
Batch-IV.}}

\author{M. Sharif \thanks{Email: msharif.math@pu.edu.pk} and G. Abbas
\\
Department of Mathematics, University of the Punjab,\\
Quaid-e-Azam Campus, Lahore-54590, Pakistan.}

\date{}
\maketitle

We deal with phantom energy accretion onto the Schwarzschild
de-Sitter black hole. The energy flux conservation, relativistic
Bernoulli equation and mass flux conservation equation are
formulated to discuss the phantom accretion. We discuss the
conditions for critical accretion. It is found that mass of the
black hole decreases due to phantom accretion. There exist two
critical points which lie in the exterior of horizons (black hole
and cosmological horizons). The results for the phantom energy
accretion onto the Schwarzschild black hole can be recovered by
taking
$\Lambda\rightarrow0$.\

Recent development in the observational cosmology reveals that our
universe is in accelerating phase. This was first confirmed by the
data of type-Ia Supernova and a large-scale structure.$^{[1-4]}$
Also, the anisotropies in cosmic microwave back ground (CMB)
radiations as observed by WMAP$^{[5-7]}$ favor the accelerating
behavior of universe. The exotic energy with negative pressure,
known as \textit{dark energy } (DE), is thought to be responsible
for this behavior of the universe. Despite of observational facts,
the nature of DE is still a challenging problem in theoretical
physics.

Different models such as quintessence $^{[8]}$, phantom $^{[9]}$,
tachyon field,$^{[10]}$, holographic $^{[11]}$ and brane-world
$^{[12]}$ models were proposed to understand the nature of DE. The
simplest form of DE is vacuum energy (cosmological constant) for
which the equation of state (EoS) parameter is $\omega=-1$. The
quintessence (dynamically evolving scalar field with negative
pressure) and phantom are in a hypothetical form of DE for which
$\omega>-1$ and $\omega<-1 $, respectively.$^{[13-15]}$ Phantom
energy violates the dominant energy condition (i.e., $\rho+p<0$,
where $\rho$ is energy density and $p$ is pressure) which results
in existence of wormholes. The expansion of the universe is
dominated by the phantom energy which diverges to approach the
future singularity (Big Rip). In this case, the phantom energy
density $\rho \rightarrow\infty$ for $t<\infty$.

Bondi$^{[16]}$ investigated the problem of matter accretion onto
the compact objects in Newtonian gravity. Michel$^{[17]}$ studied
the steady state accretion of gas onto the Schwarzschild black
hole (BH) in relativistic physics. Many researchers studied the
accretion of different forms of fluid onto the BH. Babichev {\it
et al.}$^{[18]}$ have shown that accretion of the phantom energy
onto Schwarzschild BH diminishes the BH mass. Jamil {\it et
al.}$^{[19]}$ have explored the effects of phantom accretion onto
the charged BH. They pointed out that if mass of the BH becomes
smaller (due to accretion of phantom energy) than its charge, then
the BH is converted to a naked singularity. This is the violation
of the cosmic censorship hypothesis. The same conclusion was
deduced by Babichev {\it et al.}$^{[20]}$ by studying the phantom
accretion onto the charged BH with the generalized linear EoS and
a Chaplygin gas. Madrid {\it et al.}$^{[21]}$ explored that a
Kerr-Newmann BH could be transformed to a naked singularity by the
accretion of phantom energy. In a recent study,$^{[22]}$ we have
examined the phantom accretion by the 5D charged BH.

The Schwarzschild de-Sitter (SdS) BH is a solution with vacuum
energy (cosmological constant) which helps to describe the BH
formation in the process of the universe birth$^{[23]}$. This is
important for the consideration of quantum effects near the BH in
the universe models. Many authors$^{[24-26]}$ have considered
various cosmological phenomena in the SdS spacetime. Also, our
universe is in a phase of accelerated expansion due to positive
cosmological constant (DE) and might approach to a de-Sitter phase
for $t<\infty.^{[4]}$ Martian-Moruno{\it et al.}$^{[27]}$ studied
the DE accretion on the SdS BH with FRW background. They have
found that the BH mass vanishes at big rip time. However, they do
not explore the location of critical points of accretion.

In this Letter, we investigate the phantom accretion onto a static
SdS BH by using the procedure of Jamil {\it et al.}$^{[19]}$ and
discuss the locations of the critical points of accretion. Further,
the relations between critical points and horizons are found which
were not given by Martian-Moruno {\it et al.}$^{[19]}$. The
gravitational units (i.e., the gravitational constant $G$=1 and
speed of light in vacuum $c=1$) are used. All the Latin and Greek
indices vary from 0 to 3, otherwise it will be stated.

We consider a static spherically symmetric SdS black hole given by
\begin{equation}\label{1}
ds^2=(1-\frac{2m}{r}-\frac{r^2}{a^2})dt^2-\frac{1}{(1-\frac{2m}{r}-\frac{r^2}
{a^2})}dr^2-r^2(d\theta^2+\sin\theta^2d\phi^2),
\end{equation}
where $a=\sqrt{\frac{3}{\Lambda~}}$, $m$ and $\Lambda$ are
constants. This metric has essential singularity at $r=0 $, which
is covered by the black hole horizons. Such horizons can be found
by solving $g_{00}=1-\frac{2m}{r}-\frac{r^2}{a^2}\equiv 0$ for $r$
whose positive real roots will give horizons. Using the approach
discussed in Ref.\,[23] for solving the cubic polynomial, we
explore the solution in the following three cases.

\textit{Case (i):} For $\frac{m}{a}<\frac{1}{\sqrt{27}}$, there
are three real roots of which two are positive and one is negative
(neglected). The positive and negative roots are given by
\begin{eqnarray}\label{2}
r_{\rm bh}=\frac{2a}{\sqrt{3}}\sin{\varphi},\quad r_{\rm
ch}=a(\cos{\varphi}-\frac{1}{\sqrt{3}} \sin{\varphi}),\quad
r_0=-(r_{\rm bh}+r_{\rm ch}),
\end{eqnarray}
where $\sin{3\varphi}={\sqrt{27}}\frac{m}{a}$. The subscripts bh
and ch in the above equation stand for the BH and cosmological
horizons. When $\Lambda\rightarrow 0~(a\rightarrow\infty$), we
obtain $r_{\rm bh}\rightarrow 2$m (Schwarzschild horizon) and $r_
{\rm ch}\rightarrow\infty$ (cosmological horizon does not exists).
Also, for $m\rightarrow0,~r_ {\rm bh}\rightarrow 0$ and $r_ {\rm
ch}\rightarrow a\equiv\sqrt{\frac{3}{\Lambda}}$ (de-Sitter
horizon). We would like to mention here that $r_{\rm ch}>r_{\rm
bh}$ for $0\leq\varphi<\frac{\pi}{6}$ which implies that the
exterior of the Schwarzschild BH is covered by the de-Sitter
universe and $\varphi=\frac{\pi}{6},~r_{\rm ch}=r_{\rm
bh}=\frac{a}{\sqrt{3}}$. For
$\frac{\pi}{6}<\varphi<\frac{\pi}{3}$, we have $r_{\rm ch}<r_{\rm
bh}$ which implies that the de-Sitter spacetime is the interior
structure of the Schwarzschild BH.

\textit{ Case (ii):} When $\frac{m}{a}=\frac{1}{\sqrt{27}}$, there
are three real roots of which two are positive (repeated) and one is
negative. Since negative root is neglected, so the positive roots
give unique horizon, i.e.
\begin{equation}\label{6}
r=r_{\rm bh}=r_{\rm ch}=
\frac{a}{\sqrt{3}}=\frac{1}{\sqrt{\Lambda}}.
\end{equation}

\textit{Case (iii):} For $\frac{m}{a}>\frac{1}{\sqrt{27}}$, there
are two imaginary roots and one is negative real root, hence no
horizon exists in this case.

Now we consider the phantom energy in the form of perfect fluid
whose energy-momentum tensor is
\begin{equation}\label{7}
{T_{{\mu}{\nu}}={({\rho}+p)}u_{\mu}u_{\nu}-pg_{\mu\nu}},
\end{equation}
where $\rho$ is the energy density, $p$ is the pressure and
$u^\mu=(u^t,u^r,0,0)$ is the four-vector velocity. It is mentioned
here that $u^\mu$ satisfies the normalization condition, i.e.,
$u^\mu u_\mu =1$.

The relativistic Bernoulli energy conservation equation for
accretion onto SdS black hole (using energy momentum-tensor
conservation) is given by
\begin{equation}
\label{8}
r^2u(\rho+p)\left(1-\frac{2m}{r}-\frac{r^2}{a^2}+u^2\right)^{\frac{1}{2}}=C_0,
\end{equation}
where $C_0$ is an integration constant and $u^{r}=u<0$ for inward
flow. Further, the energy flux equation can be derived by projecting
the energy-momentum conservation law on the four-velocity, i.e.,
${u_\mu T^{\mu\nu}}_{;\nu}$=0 for which Eq.(\ref{7}) leads to
\begin{equation}\label{9}
r^2u\exp
\left[\int^\rho_{\rho_\infty}\frac{d\rho'}{\rho'+p(\rho')}\right]=-C_1,
\end{equation}
where $C_1>0$ is another integration constant which is related to
the energy flux. Also, ${\rho}$ and ${\rho_\infty}$ are densities
of the phantom energy at finite and infinite $r$. From
Eqs.(\ref{8}) and (\ref{9}), we obtain
\begin{equation}\label{10}
(\rho+p)\left(1-\frac{2m}{r}-\frac{r^2}{a^2}+u^2\right)^{\frac{1}{2}}
\exp\left[-\int^\rho_{\rho_\infty}\frac{d\rho'}{\rho'+p(\rho')}\right]=C_2,
\end{equation}
where $C_2=-\frac{C_0}{C1}=\rho_\infty +p(\rho_\infty)$.

The rate of change of BH mass due to fluid accretion onto it is
given by \cite{20}
\begin{equation}\label{11}
\dot{m}=-4 \pi r^2 {T_0}^r.
\end{equation}
Using Eqs.(\ref{9}) and (\ref{10}) in the above equation yields
\begin{equation}\label{12}
\dot{m}=4 \pi C_1 [{\rho}_\infty +{p}_\infty].
\end{equation}
It is clear that $\dot{m}<0$ if $({\rho}_\infty +{p}_\infty)<0$.
Thus the accretion of phantom energy onto a BH causes to decrease
the mass of BH. Moreover, one can solve Eq.(\ref{11}) for $m$
using the EoS $p=k\rho.$ Since all $p$ and $\rho$ violating
dominant energy condition must satisfy Eq.(\ref{12}), hence it
holds in general. It is to be noted that if in Eq.(\ref{12}) the
matter contributes to the sum $({\rho}_\infty +{p}_\infty)$
instead of phantom energy, then the accretion of matter would
increase the mass of BH. Since this is not the case for matter,
there is a decrease of mass.

Now we analyze the critical points (points at which flow speed is
equal to speed of sound) during the accretion of fluid on the BH.
The fluid falls onto the BH with increasing velocity along the
particle trajectories. For any critical point $r=r_c$, we have the
following possibilities:$^{[28]}$

(1) $u^2=V^2_s$ at $r=r_c,~u^2<V^2_s$ for $r>r_c$ and $u^2>V^2_s$
for $r<r_c$. When $r\rightarrow\infty$, the flow speed is
ignorable, it is equal to the speed of sound at critical value of
$r$ while it is supersonic inside a region interior to $r_c$.

(2) $u^2>V^2_s$ and $u^2<V^2_s$ for all $r$ which are the
non-realistic cases as they describe the supersonic and subsonic
solution for all values of $r$.

(3) $u^2=V^2_s$ for all the values of $r<r_c$ or $r>r_c$. It is also
a non-physical case because it is impossible to have the same flow
speed inside and outside the critical points.

Thus the solution \textbf{1} is the only physical solution.

For the discussion of critical points of accretion onto a BH, we
follow the procedure introduced by Michel \cite{17}. The
conservation of mass flux, ${J^\mu}_{;~\mu}=0$, yields
\begin{equation}\label{13}
\rho u r^2=k,
\end{equation}
where $k$ is the constant of integration. Dividing and squaring
Eqs.(\ref{8}) and (\ref{13}), we get
\begin{equation}\label{14}
\left(\frac{\rho +p}{\rho}\right)^2 \left(
1-\frac{2m}{r}-\frac{r^2}{a^2}+u^2\right)=k_1,
\end{equation}
where $k_1=(\frac{C_0}{k})^2$ is a positive constant.
Differentiating Eqs.(\ref{13}) and (\ref{14}) and eliminating
$d\rho$, we get
\begin{equation}\label{15}
\frac{dr}{r}\left[2V^2-\frac{\frac{m}{r}+\frac{r^2}{a^2}}{1-\frac{2m}{r}
-\frac{r^2}{a^2}+u^2}\right]+\frac{du}{u}
\left[V^2-\frac{u^2}{1-\frac{2m}{r}-\frac{r^2}{a^2}+u^2}\right]=0.
\end{equation}
where $V^2=\frac{d\ln(\rho+p)}{d\ln\rho}-1$.

This equation shows that turn-around points (critical points) are
located where both the square brackets vanish. Thus
\begin{equation}\label{16}
{u_c}^2=\frac{ma^2+{r_c}^3}{2a^2r_c},\quad
{V_c}^2=\frac{ma^2+r_c^3}{2a^2r_c-3ma^2-{r_c}^3}.
\end{equation}
We see that the physically acceptable solutions of the above
equation are obtained if ${u_c}^2>0$ and ${V_c}^2>0$ implying that
\begin{eqnarray}\label{18}
{2a^2r_c-3ma^2-{r_c}^3}>0,\\
\label{19} {ma^2+{r_c}^3}>0.
\end{eqnarray}

It is worthwhile to mention here that when $a\rightarrow\infty$,
i.e., $\Lambda\rightarrow0$, the above equations reduce to the
results of accretion onto the Schwarzschild BH. The subscript $c$
is used to represent a quantity at a point where speed of flow is
equal to the speed of sound, such a point is called a critical
point. The fluid that moves towards the BH hole initially has
speed less than the speed of sound but as it comes closer to the
BH horizons, its speed may transit to a supersonic level. The
circular region around the  BH where flow speed is equal to the
speed of sound is called a sound horizon. The flow speed is
supersonic (subsonic) inside (outside) the sound horizons. Inside
the sound horizon $r<r_c$, the flow speed is supersonic but less
than the speed of light, as fluid reaches the BH horizon, the flow
speed approaches to the speed of light. After crossing the BH
horizon, it becomes greater than the speed of light.

Now we discuss the EoS and its parameter for understanding the
accretion of phantom onto the SdS BH. It is obvious that for
polytropic EoS, $p=k\rho$ and $k<0$, the speed of sound (i.e., $
{V^2}_s=\frac{\partial p}{\partial\rho}$) becomes meaningless and
fluid acts as exotic cosmic fluid that cannot be accreted onto a
BH. This problem was solved by Babichev {\it et al.}$^{[18]}$.
They introduced a generalized linear EoS, i.e.,
$p=\alpha(\rho-\rho_0)$, where $\alpha$ and $\rho_0$ are constants
and $\alpha>(<)0$ corresponds to a stable (unstable) fluid. This
EoS leads to ${V^2}_s=\alpha$. The constant $\alpha$ is related to
the EoS parameter $k$ by $k=\alpha \frac{(\rho-\rho_0)}{\rho}$.
This implies that $k<0$ corresponds to $\alpha>0$ and
$\rho<\rho_0$ and phantom energy acts as hydrodynamically stable
fluid. Thus it accretes onto the BH and diminishes its mass.

Using the procedure for solving the cubic equation mentioned in
Ref.[23], Eq.(\ref{18}) can be solved in the following cases.

\textit{Case 1:} For $\frac{m}{a}<\frac{4\sqrt{2}}{9\sqrt{3}}$,
there are three real roots of which two are positive and one is
negative (neglected). The positive roots are given by
\begin{equation}\label{19a}
r_{c_1}=\frac{2a\sqrt{2}}{\sqrt{3}}\sin\chi, \quad r_{c_2}=\sqrt{2}
a(\cos\chi-\frac{1}{\sqrt{3}}\sin\chi),
\end{equation}
where
$\sin3\chi=\frac{m}{a}\left(\frac{3}{2}\right)^{\frac{5}{2}}$. For
$0<\chi<\frac{\pi}{6}$, we have $0<r_{c_1}<r_{c_2}$, thus
$r_{c_1}$ and $r_{c_2}$ are inner and outer critical points of
flow. For $\chi=\frac{\pi}{6}$, both the critical pints coincide,
i.e., $r_{c_1}=r_{c_2}=r_c=\sqrt{\frac{2}{3}}a$ while for
$\varphi=\frac{\pi}{6},~ r_{\rm ch}=r_{\rm
bh}=r=\frac{a}{\sqrt{3}}$, hence $r_c>r$. This is not physical and
will be discussed in the following case 3. Further, taking
$0<\chi=\varphi<\frac{\pi}{6}$ and comparing Eqs.(\ref{2}) and
(\ref{19a}), we get a relation between horizons and critical
values of $r$, i.e., $r_{c_2}>r_{c_1}>r_{c_{\rm ch}}>r_{\rm
bh}>0$. This implies that curvature singularity at $r=0$ is
covered by different circular boundaries of radii
$\tilde{r}(>0)=r_{\rm bh}<r_{c_{\rm ch}}<r_{c_1}<r_{c_2}$. We
conclude from here that both the critical points lie outside the
horizons. In order to obtain the critical points, we use the
solution of Eq.(\ref{18}) given by (\ref{19a}). Substituting
Eq.(\ref{19a}) into Eq.(\ref{19}), yields
\begin{equation}
\label{24} \frac{m}{a}=\frac{16}{3}\sqrt{\frac{2}{3}}\sin
\chi(\cos^2 \chi-1),
\end{equation}
\begin{equation}
\label{25} \frac{m}{a}=\frac{2}{27}\sqrt{2}( \sqrt{3}\sin \chi -
3\cos \chi).
\end{equation}
Since $-1\leq|\cos \chi |\leq1$, Eq.(\ref{24}) implies that
$\frac{m}{a}\leq0$, which is in contradiction, hence $r_{c_1}$ is
not a physical accretion solution. Further, Eq.(\ref{25}) gives
$\frac{m}{a}>0$ for $\chi>\frac{\pi}{3}$, thus $r_{c_2}$ is a
possible critical accretion solution.

\textit{Case 2:} When $\frac{m}{a}>\frac{4\sqrt{2}}{9\sqrt{3}}$,
there is only one negative real root and other two are complex.
Consequently, this case has no physical solution and hence no
critical points.

\textit{Case 3}: For $\frac{m}{a}=\frac{4\sqrt{2}}{9\sqrt{3}}$,
there are two positive repeated roots and one is negative root.
The positive root is
\begin{equation}
\label{26} r_{c_1}=r_{c_2}=\sqrt{\frac{2}{3}}a= r_{c}.
\end{equation}
These roots are not physically critical points because
applications of these roots in Eq.(\ref{19}) lead to
$\frac{m}{a}<0$, which is in contradiction, hence this case is
also discarded.

We have devoted to study the phantom accretion onto the SdS BH.
Using the energy flux conservation, the relativistic Bernoulli
equation and the mass flux conservation equation, we formulate the
equations of motions for a steady-state spherically symmetric
phantom flow onto the SdS BH. The results reduce to the
Schwarzschild BH case $^{[17]}$ when $a\rightarrow\infty~(i.e.,
\Lambda\rightarrow0)$. The summary of the results is given as
follows:

(1) There are two horizons (BH and cosmological)and two critical
points, such that $r_{\rm bh}<r_{\rm ch}<r_{c_2}.$

(2) The particular case $r_{c_1}=r_{c_2}=r_c$ is not a physical
critical point because it gives $\frac{m}{a}<0$, which is in
contradiction.

(3) The solution $r_{c_2}$ provides a physical critical point
because it yields $\frac{m}{a}>0$ for $\chi>\frac{\pi}{6}$.

(4) Analytically, we can also determine (by using
$\chi>\frac{\pi}{6}$ in Eq.(\ref{19a})) that $r_{c_1}>r_{c_2}$, but
physically it is impossible to reverse the solution, so this case is
discarded.

(5) Physically possible critical point always lies outside the
horizons. This is according to the cases of accretion onto
Schwarzschild BH $^{[17]}$ and charged BH $^{[19]}$.

\end{document}